\documentstyle[12pt,worldsci,epsf]{article}
\def\NPB{{\em Nucl. Phys.} B}

\def\PRD{{\em Phys. Rev.} D}

\begin{document}
\title{DUAL LATTICE SIMULATIONS OF FLUX TUBES}
\author{ Martin Zach, Manfried Faber, Alexandra M\"uller, Peter Skala\\
{\em Institut f\"ur Kernphysik, Technische Universit\"at Wien, A-1040 Vienna, Austria}}
\maketitle
\setlength{\baselineskip}{2.6ex}

\vspace{0.7cm}
\begin{abstract}

Abelian gauge theories formulated on a space-time lattice can be used as a 
prototype for investigating the confinement mechanism. In $U(1)$ lattice gauge 
theory it is possible to perform a dual transformation of the path integral. 
Simulating the obtained dual theory (which corresponds to a certain limit of a 
dual Higgs model) including external sources, we perform a very accurate 
analysis of flux tubes with respect to the
dual superconductor picture.
Dual flux tube simulations are also performed in the full Abelian Higgs 
model, in order to obtain non-perturbative control over quantum and string 
fluctuations, and for a comparison to the results of dual QCD.

\end{abstract}
\vspace{0.7cm}

\section{Introduction}

The formation of colour-electric flux tubes provides an intuitive physical picture for permanent confinement of quarks within hadrons. Moreover, this picture seems to be realized within QCD, as has widely been observed in lattice simulations. The mechanism leading to flux tube formation, however, is still subject to analytical and numerical investigations.

The shortcoming of the corresponding lattice calculations, however, is the fact that regarding sufficiently long flux tubes, i.~e.~large lattice sizes, becomes a numerically very difficult task. A simpler gauge theory which also exhibits confinement is $U(1)$; it can be used as a prototype for investigating the formation of flux tubes.

Furthermore, it was realized many years ago \cite{banks} that one can perform
a duality transformation of the path integral in compact Abelian gauge
theories. In this way a new partition function is obtained which can be
regarded as a limit of the dual non-compact Abelian Higgs model
\cite{froehlich}. Besides this fact, the dually transformed theory can be used as a very efficient tool for the calculation of expectation values in the presence of
external charges \cite{pr97,np98}. The numerical advantages of such dual flux tube simulations also hold for the more general case of dual Higgs models.

In this talk we review the results obtained in dually transformed
$U(1)$ lattice gauge theory and present some first calculations in the
dual Abelian Higgs model. We will focus on one of the key questions
within the dual superconductor picture, namely whether the confinement
vacuum corresponds to a type-I or a type-II superconductor, and we will investigate the role of quantum fluctuations of the dual degrees of freedom. 

\vspace{0.2cm}
\section{Dually transformed $U(1)$ lattice gauge theory}

The expectation value of a physical observable like the electric field
or the magnetic current in the presence of external sources
(represented by Polyakov loops) can be rewritten as averaging over the
dual degrees of freedom, as has been shown in previous work \cite{pr97}. The dual representation of the Polyakov loops can be interpreted as a dual Dirac sheet connecting the electric charges.

This dual formulation opens the possibility for very accurate numerical
calculations for several reasons: Contrary to standard lattice
simulations, the confinement phase is the weakly coupled one in the
dual theory, therefore there are less quantum fluctuations. Even more
important is the fact that it is not necessary to project the
charge--anticharge state out of the vacuum: On a lattice of same size
charge pairs with arbitrary distance can be simulated with equal
accuracy, and increasing the time extent of the loops (i.~e.~of the
lattice) does not influence the quality of the result, either. 

Further advantages are the implementation of doubly charged flux tubes \cite{np98} by double Dirac strings which also yields results of equal accuracy, and the extension to periodically closed flux tubes. These ``torelons'' are free of end effects and allow for a more reliable calculation of pure flux tube properties like the string tension.

An interesting physical task is the testing of the validity of a dual
London equation for $U(1)$ flux tubes \cite{haymaker}. In the dual
formulation we were able to investigate rather long flux tube
lengths up to 20 lattice spacings. Our results \cite{pr97} showed that
the agreement between the $U(1)$ data and the predictions of a classical
model of Maxwell and London equations are very good for small charge
distances, but there is absolutely no agreement at large distances
($>10$ lattice spacings). A possible interpretation within the dual
superconductor picture is that the confining $U(1)$ behaves as an
effective type-I superconductor, rather than an extreme type-II as an
exact validity of the dual London equation would suggest. In fact we got
support for this conjecture also from a completely
independent investigation: simulating doubly charged flux tubes. We
found \cite{np98} that in four-dimensional $U(1)$ there is an
attractive interaction between flux tubes for the coupling 
$\beta$ approaching the phase transition. 

Thus one might look at $U(1)$ lattice gauge theory from two points of
view: From a ``microscopical'' it can be regarded as a double limit of a dual Higgs model with both vanishing London penetration length and vanishing coherence length. From a ``macroscopical'' point of view it looks like a classical dual type-I superconductor. This peculiarity is related to the fluctuations in the dual theory: The observed attraction between flux tubes was shown to be a purely ``quantum-mechanical'' effect of the dually transformed $U(1)$ theory \cite{np98}.

\vspace{0.2cm}
\section{The dual Abelian Higgs model} 

The action of the full Abelian Higgs model including a dual Dirac
string (denoted with $^*n$) can be written in the form
\begin{equation}\label{s_higgs}
S = \beta \! \sum_{\it plaqu.} \! \left({d}\, {}^*\theta + 2 \pi {}^*
n\right)^2 \; - \; 2 \gamma \sum_{\it links} |\overline{{}^*\Phi}|^2
\cos \left(d\,{}^*\chi - \,{}^*\theta \right) \;+\; \kappa \sum_{\it sites}\;V(^*\Phi),
\end{equation}
where $^*\theta$ is the dual gauge field and $^*\Phi$ the dual Higgs
field, located on the sites of the dual lattice. $V(^*\Phi)$ is the
symmetry-breaking Higgs potential with its minimum at $^*\Phi=1$. For
the squared covariant derivative of the Higgs current the compact
formulation has been chosen, because this allows for fluctuations of
the fluxoid string \cite{pr97}. In the classical model $\beta/\gamma$ is the squared London
penetration depth, and $\gamma/\kappa$ the squared coherence
length. The double limit $\gamma \to \infty$, $\kappa \to \infty$
exactly corresponds to the dually transformed $U(1)$ theory discussed
in the last section.

If we compare this model to the parameters of dual QCD \cite{baker}
after an Abelian ansatz, we find the correspondence
\begin{equation}
\beta = \frac{4}{3} \; \frac{1}{g^2}, \quad
\gamma = 8 \;B_0^2 \;a^2, \quad
\kappa = \frac{100}{3} \;\lambda \;B_0^4 \;a^4,
\end{equation}
where $g$ is the (chromo-)magnetic coupling, $B_0$ the location of the
minimum and
$\lambda$ the strength of the corresponding continuum 
Higgs potential, and $a$ is the lattice spacing. 
These parameters can be chosen to reproduce very well the infrared
behaviour of QCD. The classical flux tube solutions of the Higgs model on
sufficiently fine lattices give very good agreement with 
corresponding continuum calculations \cite{nora}. It is now an 
interesting question how quantum and string fluctuations (of the
fluxoid string) do effect the physical quantities, e.~g.~the string
tension. 

\begin{figure}[t]
\begin{centering}
\centerline{\epsfbox{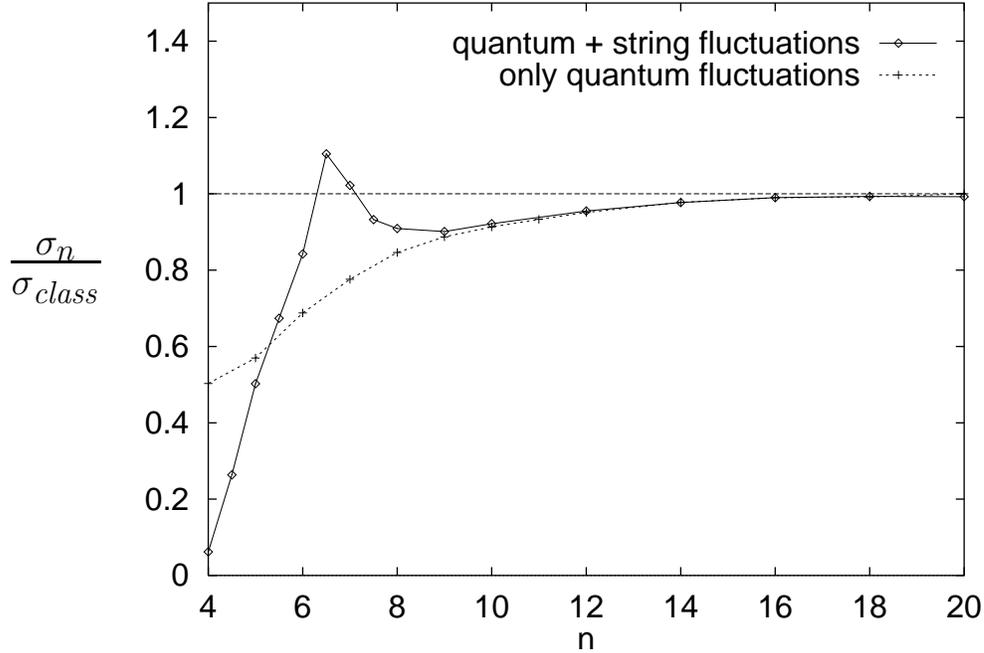}}
\begin{picture}(0,0)(0,0)
\put(40,160){\makebox(0,0){\LARGE $\frac{\sigma_n}{\sigma_{\it {class}}}$}}
\end{picture}
\end{centering}
\vspace{-0.3cm}
\caption{The changing of the string tension $\sigma$
with respect to the classical value $\sigma_{\it {class}}$ under the
influence of quantum fluctuations (dotted line), and of both quantum
and string fluctuations (solid line). This calculation has been
performed for a torelon on a $8^4$ lattice.}
\label{sigma_n}
\end{figure}

For the set of parameters which yields the desired classical solution,
the system is not in the Higgs phase any more on a reasonably fine lattice, if fluctuations are
turned on. For this reason we introduce a ``quantum scale'' $n$, by
substituting $\exp(-S)$ in the path integral by $\exp(-n\,S)$, which
allows for a smooth interpolation between the classical limit
($n=\infty$) and an ordinary simulation ($n=1$). The changing of the
string tension as a function of $n$ is depicted in Fig.~\ref{sigma_n}
for two cases: Considering only quantum fluctuations but keeping the
physical fluxoid string fixed, and also considering string
fluctuations, which leads to an increase of the energy per length
(until the system undergoes a phase transition to the Coulomb phase).
%\vspace{0.3cm}

Physical arguments suggest that the dual superconductor should be
approximately at
the borderline between type-I and type-II. This is fulfilled for the chosen
classical solution. Investigating the influence of fluctuations on an
effective Landau-Ginzburg parameter (defined by the ratio of the
individual contributions to the string tension) showed us that the
system is pushed towards type-I behaviour by fluctuations.

\vspace{0.2cm}
\section{Conclusions}

By dual simulations of flux tubes one can address a lot of interesting
questions concerning the confinement mechanism. Here we focussed on
the discussion whether the confinement vacuum is a dual type-I or
type-II superconductor, and on the meaning of fluctuations in dual
theories. Similarly to the behaviour of $U(1)$ in the vicinity of the
phase transition, these fluctuations can produce an effective type-I
superconductor also in the full Abelian Higgs model.

Of course, it is another interesting question, which set of (bare)
parameters has to be used to describe the same physics as with the
classical solution of the Abelian Higgs model. For this task, however,
it will be necessary to improve the control over string fluctuations
in our simulations.

\vspace{0.2cm}
\section*{Acknowledgments}
This work was supported by the Fonds zur F\"orderung der wissenschaftlichen Forschung, Project No.~P12495-TPH.

\end{document}